\documentclass[prb,superscriptaddress,twocolumn,longbibliography]{revtex4-1}
\usepackage{times,amsmath}
\usepackage{epsfig}
\usepackage{color}
\usepackage{longtable}
\usepackage{ulem}
\usepackage{graphicx}
\usepackage{dcolumn}
\usepackage{bm}
\usepackage{bookmark}
\usepackage{tabularx}
\usepackage{hyperref}
\usepackage{multirow}
\hypersetup{colorlinks=true, citecolor=blue, filecolor=blue, linkcolor=blue, urlcolor=blue}

\begin{document}

\title{Semiconducting electrides derived from Sodalite: A first-principles study}

\author{Chang Liu}
\affiliation{Department of Physics and Astronomy, University of Nevada, Las Vegas, NV 89154, USA}

\author{Mahfuza Mukta}
\affiliation{Department of Mechanical Engineering and Engineering Science, University of North Carolina at Charlotte, Charlotte, NC 28223, USA}

\author{Byungkyun Kang}
\email{bkang@udel.edu}
\affiliation{College of Arts and Sciences, University of Delaware, Newark, Delaware 19716, USA}

\author{Qiang Zhu}
\email{qzhu8@uncc.edu}
\affiliation{Department of Mechanical Engineering and Engineering Science, University of North Carolina at Charlotte, Charlotte, NC 28223, USA}
\date{\today}

\begin{abstract}
Electrides are ionic crystals with electrons acting as anions occupying well-defined lattice sites. These exotic materials have attracted considerable attention in recent years for potential applications in catalysis, rechargeable batteries, and display technology. Among this class of materials, electride semiconductors can further expand the horizon of potential applications due to the presence of a band gap. However, there are only limited reports on semiconducting electrides, hindering the understanding of their physical and chemical properties. In a recent work, we initiated an approach to derive potential electrides via selective removal of symmetric Wyckoff sites of anions from existing complex minerals. Herein, we present a follow-up effort to design the semiconducting electrides from parental complex sodalites. Among four candidate compounds, we found that a cubic Ca$_4$Al$_6$O$_{12}$ structure with the $I$-43$m$ space group symmetry exhibits perfect electron localization at the sodalite cages, with a narrow electronic band gap of 1.2 eV, making it suitable for use in photocatalysis. Analysis of the electronic structures reveals that a lower electronegativity of surrounding cations drives greater electron localization and promotes the formation of an electride band near the Fermi level. Our work proposes an alternative approach for designing new semiconducting electrides under ambient conditions and offers guidelines for further experimental exploration.

\end{abstract}

\vskip 300 pt
\maketitle

\section{Introduction}
Electrides are exotic ionic materials in which electrons occupy the well-defined lattice sites and serve the role of anions. These materials were originally discovered by Dye and coworkers in organic salts in the 1980s, however, their high sensitivity to heat and oxidation limits their practical application \cite{Ellaboudy-1983-JACS, Dye-Science-2003}. Subsequent efforts have been directed towards achieving stable electrides, aiming to explore chemical and physical properties arising from nearly-free electrons and their corresponding geometric topologies. In 2003, the first thermally stable electride Ca$_6$Al$_7$O$_{16}$ (C12A7:2$e^-$) was synthesized from the inorganic mineral mayenite (12CaO$\cdot$7Al$_2$O$_3$) \cite{Matsuishi-Science-2003}. The resulting electride, with excess electrons confined in zero dimensional (0D) cages, exhibits excellent thermal stability and low reactivity with air. This discovery marks a milestone in the research on electrides, and spurred numerous new efforts to design other novel inorganic electrides. To date, electrides have been identified in various inorganic compounds exhibiting different configurations of confined electrons \cite{burton2018high, ZHU20191293}, including 0D cavities \cite{Matsuishi-Science-2003, Ca3Pb-2018, wang2018ternary}, 1D chains \cite{Wang-JACS-2017, Zhang-JPCL-2015, Lu-JACS-2016, Lu-PRB-2018, chanhom2019sr3crn3}, 2D planes \cite{Lee-Nature-2013, Tada-IC-2014, Inoshita-PRX-2014, druffel2017electrons, Wan2018}, and 3D configurations \cite{tada-2017}. These diverse electronic configurations produce unique properties, including high electron concentration \cite{Lee-Nature-2013, Kim2006}, anti-ferromagnetism \cite{Lu-PRB-2018},  thermionic electron emission at low temperature \cite{Lee-Nature-2013, toda2004field}, high density of active sites \cite{wu2017tiered}, and low work function \cite{Toda-AM-2007}. These exceptional properties make electrides promising candidates for applications in catalysis \cite{Kitano-NChem-2012, Kitano-NC-2015}, energy storage \cite{hu20152d, kocabas2018determination}, and electronics \cite{Park-JACS-2017, Park-JACS-2017-tune}.

While most currently reported electrides are semi-metals, an electride with semiconducting band structure has the potential to further expand the horizon of potential applications of such electrides as infrared photodetector \cite{mcrae2022sc2c}. Previous theoretical study proposed a transition in Ca$_2$N from 2D conducting electride to a 0D semiconducting electride state under high pressure \cite{zhang2017pressure}. This discovery was supported by electrical resistance measurements conducted in a follow-up experimental study \cite{tang2018metal}. To further investigate physical and chemical properties and explore possible applications of semiconducting electrides, stable structures at ambient conditions still need to be developed \cite{wang2023transition}. In 2022, a novel inorganic electride, Sc$_2$C, was discovered as the first 2D electride exhibiting semiconducting behavior \cite{mcrae2022sc2c}. Its small band gap and good conductivity enable applications as a battery electrode or in IR photodetectors. Recent theoretical study identified Mg$_2$N in $R$3$m$ symmetry as a 0D electride with a semiconducting band structure \cite{PhysRevApplied.19.034014}. But research on this novel class of electrides is still in its infancy. Consequently,  discovering new semiconducting electrides with different confined electron configurations stabilized at ambient conditions is essential to understand related properties and new electride design towards potential applications.

\begin{figure*}[htbp]
\centering
\includegraphics[width=0.95 \textwidth]{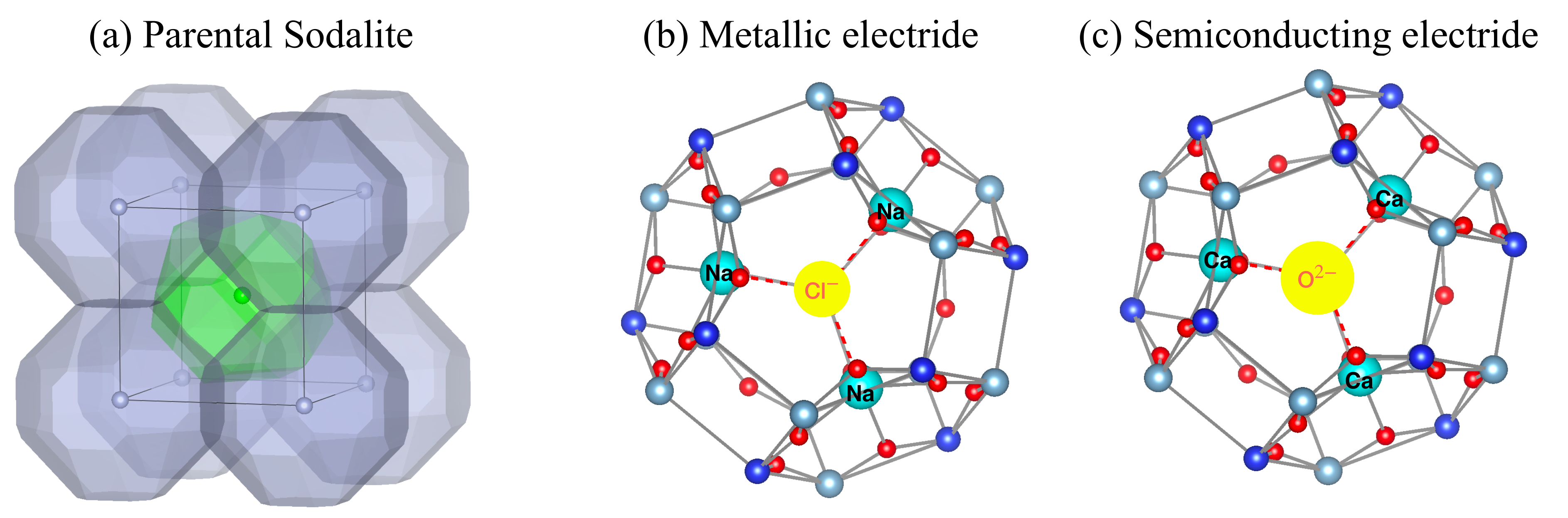}
\caption{\label{Fig1} The schematic illustration of electride properties } 
\vspace{-3mm}
\end{figure*}

Recently, we introduced an alternative method to achieve electride states by removing high-symmetry Wyckoff sites of anions from existing sodalite compounds \cite{kang2023first}. The resulting sodalite electrides are expected to exhibit greater thermal stability due to their intricate structural framework, thereby holding promise for practical applications. In our previous work, we primarily focused on halides, where only one electron can be accommodated per cage in the resulting sodalite framework (see Fig. \ref{Fig1}). Consequently, the localized cage electrons can lead to either a half-metallic state or a Mott insulator from the perspective of electron counts \cite{kang2023first}. According to the electron counting rule, it is likely that two localized electrons in a cage of the crystal structure may form a distinct electronic band with full occupation, leading to a semiconducting or insulating band structure. Although this simple model may be complicated by the interplay between the cage electrons and their surrounding cations, we hypothesize that this is a viable approach to design new electride materials with a nonzero band gap. Therefore, we have decided to further explore this avenue by investigating sodalite compounds that contain high-symmetry Wyckoff sites of anions with a -2 formal charge.

In this work, we applied a screening strategy similar to our previous study \cite{kang2023first} on existing multi-component sodalites and identified four candidate parental structures characterized by the presence of chalcogen anions in high-symmetry Wyckoff sites. These candidates include Ca$_4$Al$_6$O$_{13}$, Zn$_4$B$_6$O$_{13}$, Cd$_4$Al$_6$SO$_{12}$, and Be$_3$Cd$_4$Si$_3$SeO$_{12}$. We initially examined the lattice dynamical properties of these structures after removing the anionic site at the center of the unit cell. Theoretical results indicate that the generated compounds are dynamically stable under ambient conditions. Subsequently, we conducted electronic structure calculations on these derivative compounds to verify the presence of electride states. Interestingly, we found that these sodalite compounds, with various metal cations adjacent to the electride site, provide an ideal platform for exploring systematic trends. Our further analysis of the evolution of electronic properties with the adjacent cations revealed that a lower electronegativity of the adjacent cation can stabilize the electride states and generate a distinct electride band around the Fermi level.

\section{Computational Methods}

Following a query of the Materials Project \cite{MP-2013} using our in-house symmetry analysis toolbox, \texttt{PyXtal} \cite{pyxtal}, we selected four complex sodalite structures in cubic symmetry, Ca$_4$Al$_6$O$_{13}$, Zn$_4$B$_6$O$_{13}$, Cd$_4$Al$_6$SO$_{12}$, in $I$-43$m$ space group and Cd$_4$Be$_3$Si$_3$SeO$_{12}$ in $P$43$n$ space group as parent structures to create the electride phases. 
For each structure, the electride configuration was generated by removing the anions at the center of the unit cell.
The structures were then fully relaxed to obtain optimal cell parameters, and their structural stability was checked using phonon calculations. All calculations were performed using the projector augmented wave (PAW) method \cite{PAW-PRB-1994}, implemented in the \texttt{VASP} code \cite{Vasp-PRB-1996} within the framework of density functional theory (DFT). The generalized gradient approximation (GGA) with the Perdew, Burke, and Ernzerhof (PBE) functional \cite{PBE-PRL-1996} was adopted. For geometric relaxation, we used a unit cell containing two formula units and a 3 $\times$ 3 $\times$ 3 $\Gamma$-centered k-point grid was adopted. To simulate electronic properties, a primitive cell with a dense mesh of 8 $\times$ 8 $\times$ 8 k-point grid was utilized. The cutoff energy for all calculations was set to 520 eV, achieving convergence for energy around 1 meV per atom and for forces within 0.01 eV/\AA. For phonon calculations, the optimized structures for electrides were used to construct a 2 $\times$ 2 $\times$ 2 supercell (containing 176 atoms), employing a single $\Gamma$ point for Brillouin zone sampling. Phonon density of states were calculated using force constants obtained via the finite displacement method implemented in the \texttt{Phonopy} code \cite{Togo-PRB-2008}.

\begin{figure*}[htbp]
\centering
\includegraphics[width=0.96 \textwidth]{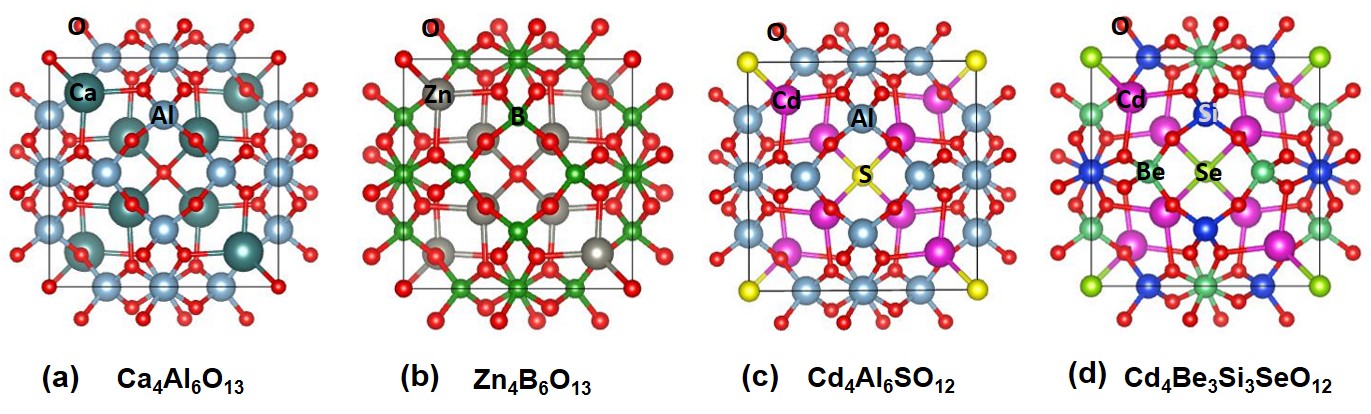}
\caption{\label{Fig2} Crystal structures of (a) Ca$_4$Al$_6$O$_{13}$, (b) Zn$_4$B$_6$O$_{13}$, (c) Cd$_4$Al$_6$SO$_{12}$, (d) Cd$_4$Be$_3$Si$_3$SeO$_{12}$. The high symmetry cage anion sites are located at the body center position in each structure.} 
\end{figure*}

\section{Results and Discussions}

\subsection{Crystal Structures}

In this work, all four candidate parental oxides have already been observed in experiment \cite{ponomarev1971crystal, smith1980redetermination, brenchley1993synthese, dann1996synthesis}. 
As shown in Fig. \ref{Fig2}, each unit cell contains two formula units. According to X-ray diffraction analysis \cite{ponomarev1971crystal}, Ca$^{2+}$ is bonded to four O$^{2-}$ atoms to form distorted CaO$_4$ trigonal pyramids that share corners with six equivalent AlO$_4$ tetrahedra and corners with three equivalent CaO$_4$ trigonal pyramids. Other materials, including Zn$_4$B$_4$O$_{13}$, Cd$_4$Al$_4$SO$_{12}$, and Cd$_4$Be$_3$Si$_3$SeO$_{12}$ show the similar packing behavior. Table \ref{tab_pro} summarizes the calculated cell parameters, which align well with experimental data, thus confirming the choice of DFT calculation parameters.

\begin{table}[ht]
\caption{The crystallographic data and band gaps of four parent sodalites and their derivative electride candidate structures.}\label{tab_pro}
\begin{tabular}{lcllc}
\hline\hline
System    & Space Group & \multicolumn{2}{l}{~~~~~~$a$ (\AA)~~~~~~} & ~~Band Gap~~  \\
          &             & ~~DFT~~                & ~~Expt.~~     &    (eV) \\\hline
\multicolumn{5}{l}{Sodalites} \\\hline                
Ca$_4$Al$_6$O$_{13}$ & $I$-43$m$  & ~~8.87  & ~~8.86\cite{ponomarev1971crystal}  & 3.9     \\
Zn$_4$B$_6$O$_{13}$  & $I$-43$m$  & ~~7.55  & ~~7.47\cite{smith1980redetermination}  & 3.4     \\
Cd$_4$Al$_6$SO$_{12}$ & $I$-43$m$ & ~~8.95 & ~~8.82 \cite{brenchley1993synthese}&2.7\\     
Cd$_4$Be$_3$Si$_3$SeO$_{12}$ & $P$43$n$ & ~~8.62 & ~~8.49 \cite{dann1996synthesis} & 2.9\\\hline
\multicolumn{5}{l}{Electride candidates}  \\\hline
Ca$_4$Al$_6$O$_{12}$ & $I$-43$m$  & ~~9.06  &  & 1.2\\
Zn$_4$B$_6$O$_{12}$  & $I$-43$m$  & ~~7.41  &  & 3.6  \\
Cd$_4$Al$_6$O$_{12}$ & $I$-43$m$ & ~~8.64  &  & 2.7 \\
Cd$_4$Be$_3$Si$_3$O$_{12}$ & $P$43$n$ & ~~8.22 && 3.0\\\hline\hline            
\end{tabular}
\end{table}

\begin{figure*}[htbp]
\centering
\includegraphics[width=0.99 \textwidth]{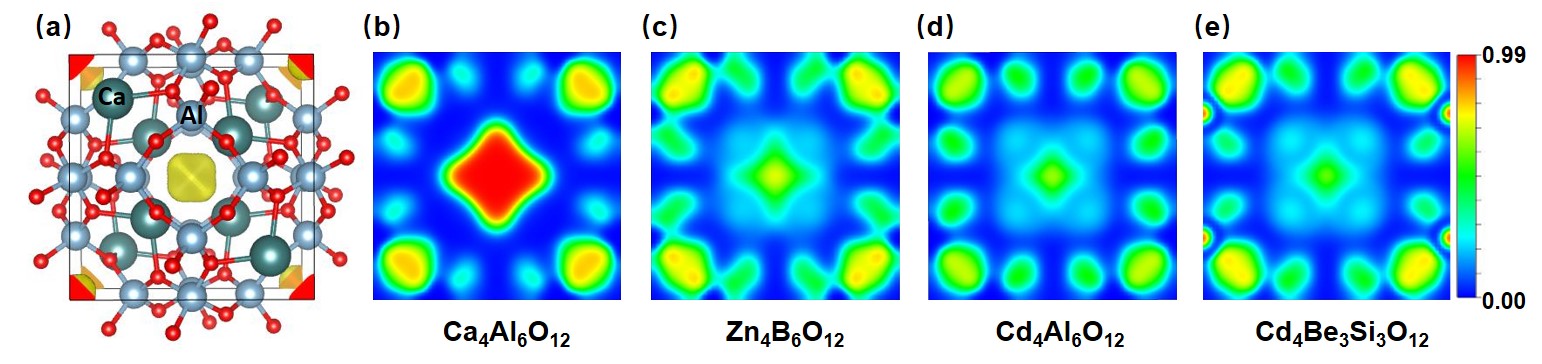}
\caption{\label{Fig3} Calculated electron localization function (ELF) after the removal of O$^{2-}$. (a) the 3D ELF isosurface ELF of Ca$_4$Al$_6$O$_{12}$. (b-e) 2D ELF of Ca$_4$Al$_6$O$_{12}$, Zn$_4$B$_6$O$_{12}$, Cd$_4$Al$_6$O$_{12}$, Cd$_4$Be$_3$Si$_3$O$_{12}$ in the (200) plane.} 
\end{figure*}

For each structure, we removed the cage anions at the body center of the unit cells to generate candidate electride structures. Specifically, we removed O$^{2-}$ for Ca$_4$Al$_6$O$_{13}$ and Zn$_4$B$_6$O$_{13}$, S$^{2-}$ for Cd$_4$Al$_6$SO$_{12}$, and Se$^{2-}$ for Cd$_4$Be$_3$SeO$_{12}$. Correspondingly, the resulting candidate electride structures consist cages surrounded by four Ca$^{2+}$ in Ca$_4$Al$_6$O$_{12}$, four Zn$^{2+}$ in Zn$_4$B$_6$O$_{12}$, four Cd$^{2+}$ in both Cd$_4$Al$_6$O$_{12}$ and Cd$_4$Be$_3$Si$_3$O$_{12}$. Comparing the cell parameters between the parental and anion-removal phases, we found that most of the systems undergo a lattice shrink after the removal of anions (see Table \ref{tab_pro}). However, the Ca$_4$Al$_6$O$_{12}$ is an exception. After the removal of O$^{2-}$, the volume unexpectedly increases, suggesting that new types of interactions may form due to the localized cage electrons.

Additionally, we computed the phonon spectrum for each structure and found that none exhibit imaginary frequencies (see Fig. S1 in the supplementary materials). This indicates that all structures remain dynamically stable after the removal of cage anions, allowing them to persist as long as they are synthesized in experimental conditions.

\subsection{Spatial Electron Localization}

To characterize the electrides state, we calculated the electron localization function (ELF), which measures the degree of electron's spatial localization compared to a reference electron with the same spin. The ELF value, bounded between 0 and 1, reflects the likelihood of finding an electron in the neighborhood of a reference electron located at a given point in which ELF = 1 corresponding to perfect localization and ELF = 0.5 corresponding to free electron gas \cite{ELF}. 

As shown in Fig. \ref{Fig3}, we observed strong ELF values at the body center cages after the removal of the cage anions from the parent structures. These results suggest that the electrons remain localized at the sodalite cages. Among them, Ca$_4$Al$_6$O$_{12}$, possesses an ELF value of 0.99 at the body center, suggesting the strongest electron localization. On the other hands, the ELF values at the cage centers for Zn$_4$B$_6$O$_{12}$, Cd$_4$Al$_6$O$_{12}$ and Cd$_4$Be$_3$Si$_3$O$_{12}$ are 0.72, 0.65 and 0.59, respectively. The trend of the ELF values has strong correlation with the electronegativity of the surrounding metal cations. Ca has an electronegativity value of 1.00, whereas Zn has a value of 1.65 and Cd of 1.69 \cite{allen1989electronegativity}. Due to its lower electronegativity compared to Zn$^{2+}$ and Cd$^{2+}$ cations, Ca$^{2+}$ is expected to have a weaker attraction to the outer core electrons, consequently leading to the greater electron localization at the body center cages in Ca$_4$Al$_6$O$_{12}$.

\begin{figure*}[htbp]
\includegraphics[width=0.99 \textwidth]{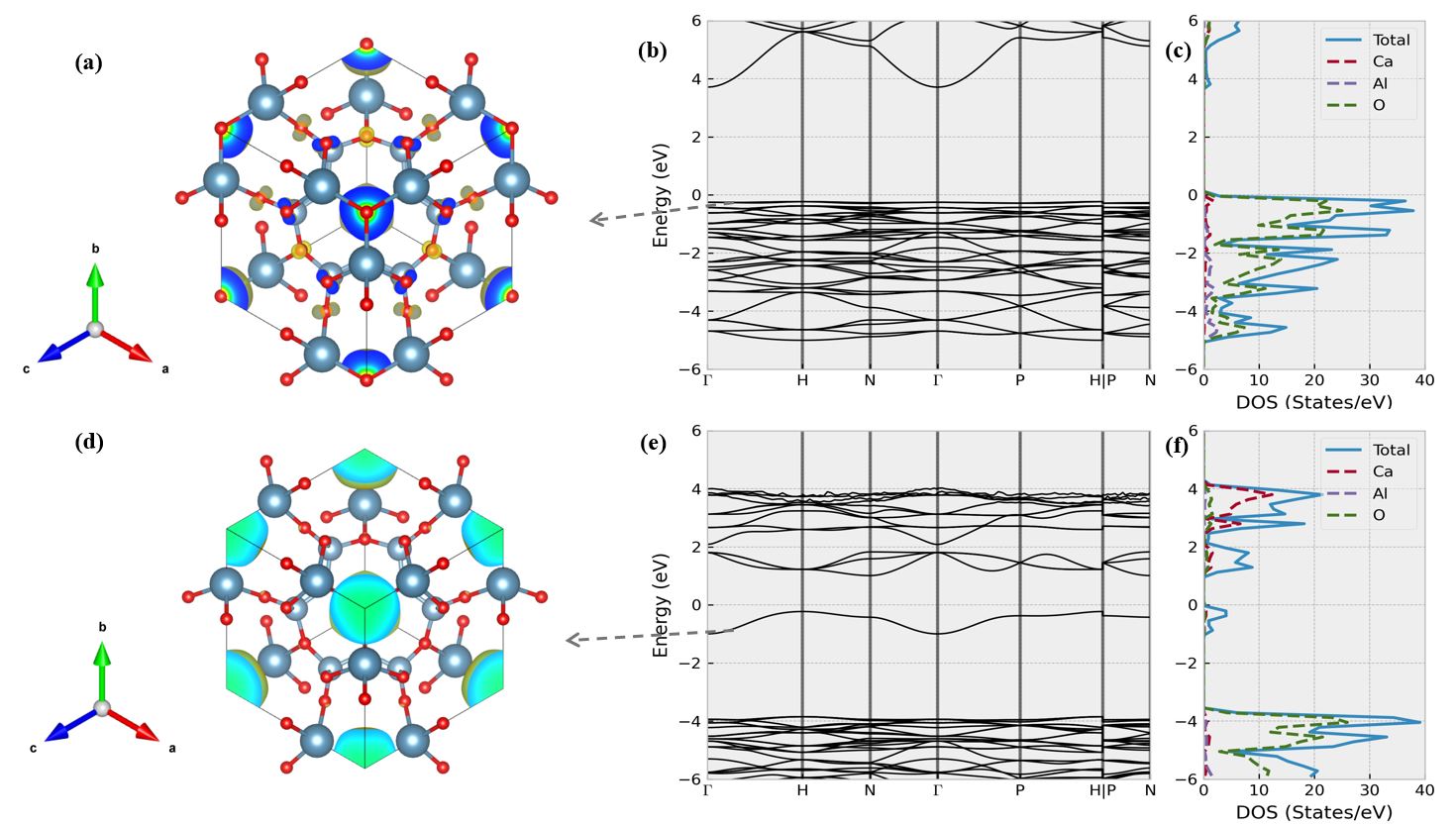}
\caption{\label{Fig4} Electronic structures of Ca$_4$Al$_6$O$_{13}$ and Ca$_4$Al$_6$O$_{12}$. (a) displays the isosurface of the decomposed charge density of Ca$_4$Al$_6$O$_{13}$'s highest valence band, whereas (b) and (c) plot its band dispersion and density of states (DOS) at an extended energy range around the Fermi level. As a comparison, (d)-(f) show the isosurface of the decomposed charge density of Ca$_4$Al$_6$O$_{12}$'s highest conduction band and its corresponding band structure and DOS plots. The isosurface value is set as 0.003 e/bohr$^3$ in (a) and (d).} 
\end{figure*}

\subsection{Electronic Band Structures}
In addition to the electron's spatial localization, we are also interested in understanding their electronic band structures and the energy levels of those localized cage electrons. As shown in Fig. \ref{Fig4}b, the parent structure Ca$_4$Al$_6$O$_{13}$ is a semiconductor with a band gap of 3.9 eV, and the highest valence band (HVB) near the Fermi level are very flat. The associated DOS plot (Fig. \ref{Fig4}c) suggests that HVB is mainly contributed by the O orbitals, while the conduction bands comprise the hybridization of oxygen and metal atoms. More precisely, we can clearly see that the decomposed charge density of HVB is mainly featured by the electrons around the O$^{2-}$ anions at the cage center, as shown in Fig. \ref{Fig4}a. Overall, the HVB in Ca$_4$Al$_6$O$_{13}$ has an energy close to other valance bands.

After the removal of cage O$^{2-}$ anions, the resulting structure maintains a semiconducting band structure with a much narrower band gap of 1.2 eV. As shown in both Fig.\ref{Fig3}d and e, the Ca$_4$Al$_6$O$_{12}$ has a HVB clearly separated from other valence bands in the systems. Remarkably, the computed partial charge configuration in Fig. \ref{Fig4}d, unlike what is shown in Fig. \ref{Fig4}a,  suggests that this band solely correspond to electrons localized at cage centers. 

This scenario can be simply understood from the electron counting rule as we presented earlier. When O is present at the cage center, each O has the capability to attract excess electrons to from four neighboring Ca atoms, leading to a charge transfer from Ca to O and forming the ionic bonding between O$^{2-}$ and Ca$^{2+}$. This ionic bonding can also helps to lower the valence band energy. When the cage O atoms are removed, the excess electrons from the Ca atoms cannot be redistributed to any surrounding cations. As a result, the localized cage electrons behaves like a nucleus-free anions to stabilize the crystal structure. If there is only one electron in each cage, the system should form a partially occupied band as discussed in our previous work \cite{kang2023first}. Given that two electrons are present in each cage of Ca$_4$Al$_6$O$_{12}$, these electrons forms a fully occupied electronic band. And the whole structure remains semiconducting. However, this band is expected to possess a higher energy as compared to the counterpart HVB in Ca$_4$Al$_6$O$_{13}$. Notably, the repulsion between localized cage electrons and surrounding Ca$^{2+}$ in Ca$_4$Al$_6$O$_{12}$ also tend to expand the [Ca]$_4$ tetrahedra and thus leads a larger volume as compared to the parental phase. Therefore, the band gap in Ca$_4$Al$_6$O$_{12}$ becomes smaller. And we can safely conclude that Ca$_4$Al$_6$O$_{12}$ is a typical electride feature with a semiconducting band gap. To our knowledge, this is the first report of non-layered electride phase that can be stabilized at ambient pressure conditions. 

\begin{figure*}[htbp]
\centering
\includegraphics[width=0.95 \textwidth]{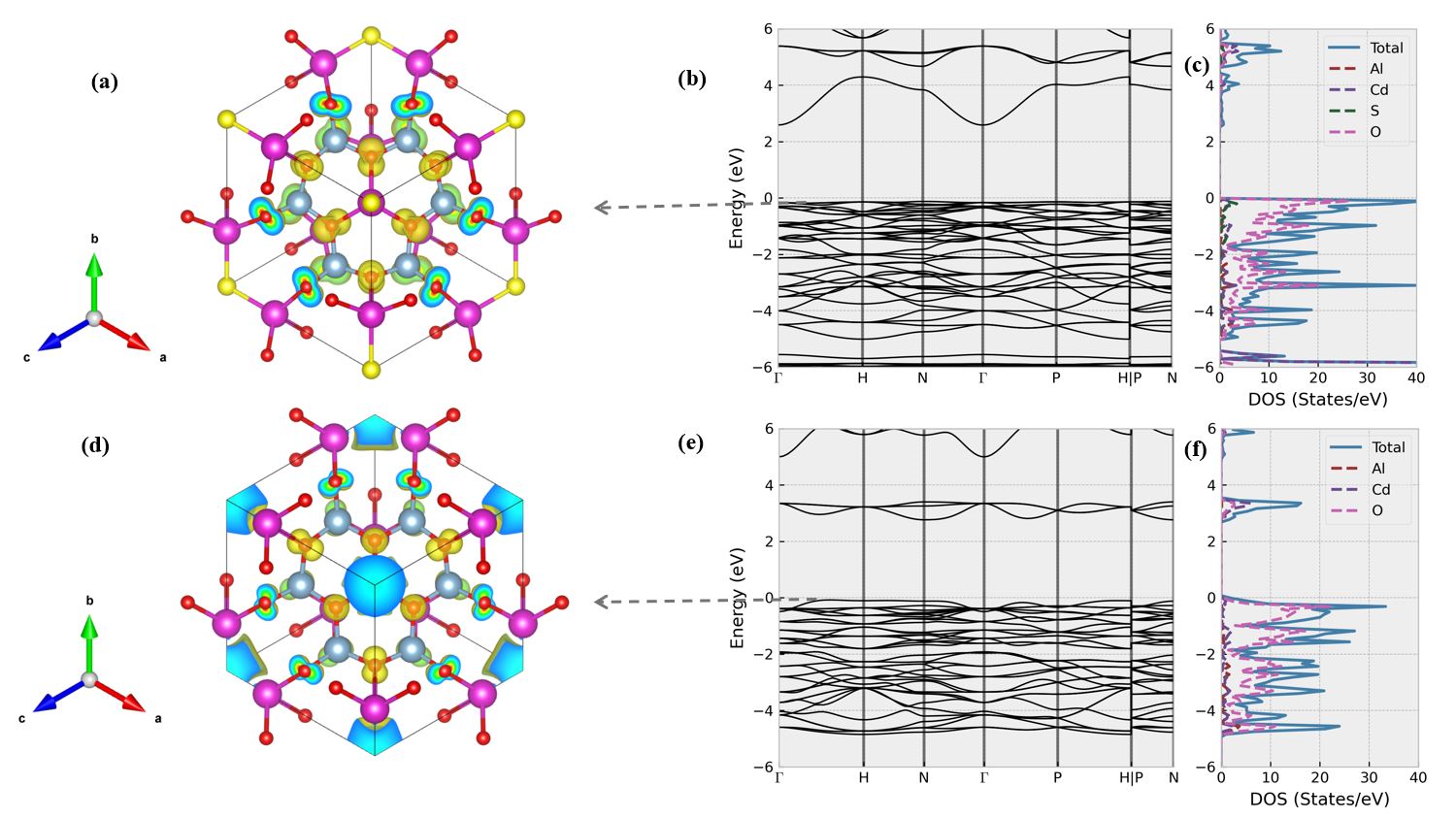}
\caption{\label{Fig5} Electronic structures of the Cd$_4$Al$_6$SO$_{12}$ and Cd$_4$Al$_6$O$_{12}$. (a) displays the decomposed charge density of Cd$_4$Al$_6$SO$_{12}$'s highest valence band, whereas (b) and (c) plot its band dispersion and DOS at the extended range. As a comparison, (d)-(f) show the isosurface of the decomposed charge density of Cd$_4$Al$_6$O$_{12}$'s highest valence band, as well as its corresponding band structure and DOS plots. The isosurface value is set as 0.003 e/bohr$^3$ in (a) and (d).} 
\end{figure*}

While the case of Ca$_4$Al$_6$O$_{12}$ demonstrate the possibility to achieve a energy band fully occupied by interstitial electrons near the Fermi level, the interplay with the local chemical environment may complicate this scenario. As shown in Fig. \ref{Fig5}, the parent system Cd$_4$Al$_6$SO$_{12}$ is a semiconductor with a band gap of 2.7 eV. The top valence bands near the Fermi level of Cd$_4$Al$_6$SO$_{12}$ mainly come from O, S and Cd atoms. Upon removing the S atoms, we can find several high energy conduction bands drop off, similar to those in Ca$_4$Al$_6$O$_{12}$. However, the band gap of the Cd$_4$Al$_6$O$_{12}$ does not vary significantly from that of the parent structure. This is also evidenced by the visualization of partial charge density of the valence band closest to the Fermi level (see Fig. \ref{Fig5}d). While there exist localized electrons around the cage center, we also find a significant portion of electrons around the neighboring O atoms (that are similar to the parental compound). Therefore, we can better interpret this band reflects an interaction between Cd$_4$ cluster cations and neighboring O anions. Compared to cage electrons in Ca$_4$Al$_6$O$_{12}$, the cage electrons in Cd$_4$Al$_6$O$_{12}$ are tightly bounded by the Cd nucleus. Therefore the charge redistribution after the removal of S atoms in Cd$_4$Al$_6$SO$_{13}$ fails to generate an isolated energy band fully occupied by interstitial electrons near the Fermi level. We also observed a similar trend in Cd$_4$Be$_3$Si$_3$SeO$_{12}$ (see Fig. S2 in the supplementary materials). In this system, the corresponding partial charge density plot in Cd$_4$Be$_3$Si$_3$O$_{12}$ reveals that O contribute more portions as compared to the cage electrons, suggesting that the substitution of Be and Si can systematically shift the charge redistribution.

\begin{figure*}[htbp]
\centering
\includegraphics[width=0.95 \textwidth]{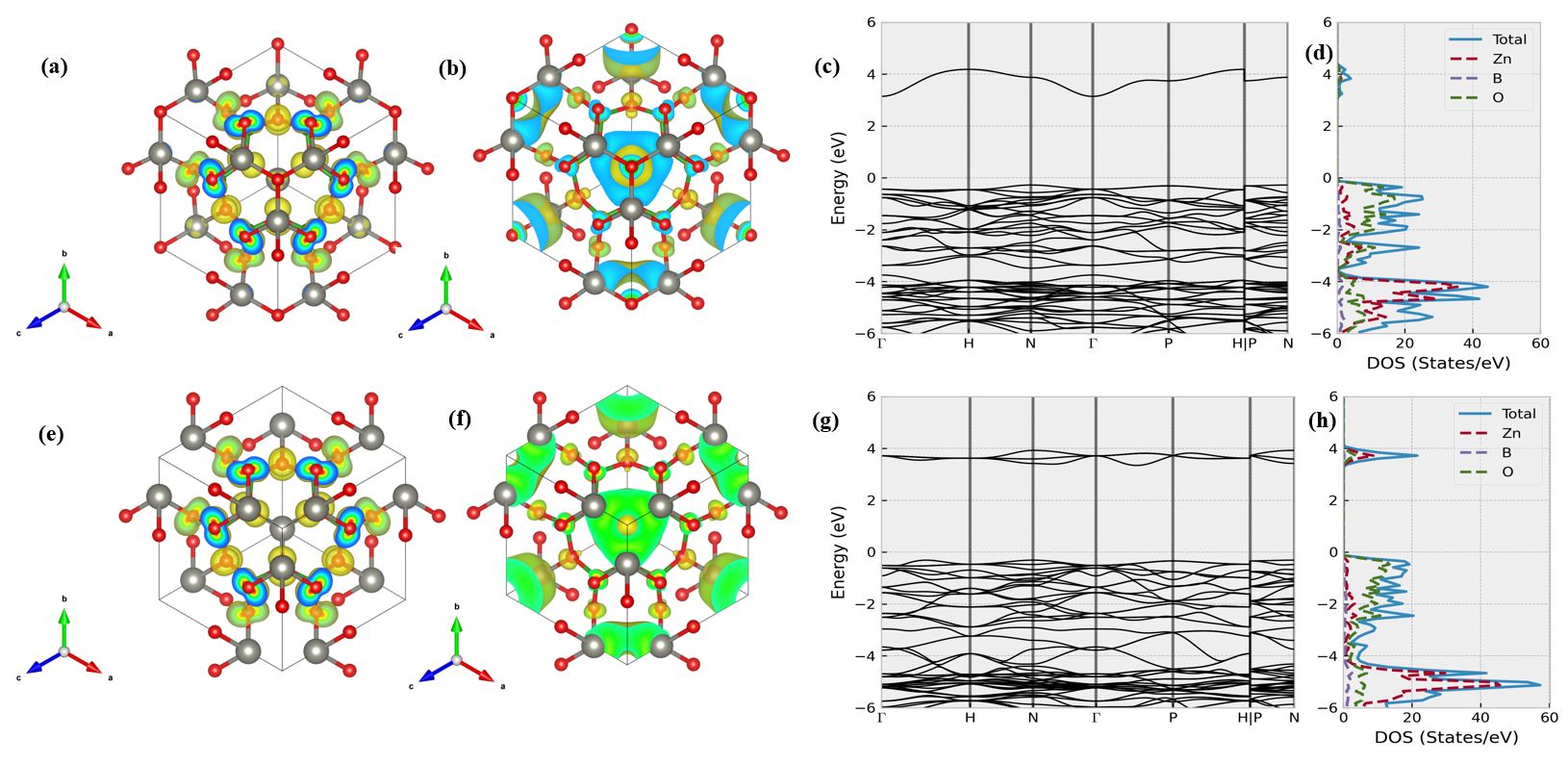}
\caption{\label{Fig6} Electronic structures of the Zn$_4$B$_6$O$_{13}$ and Zn$_4$B$_6$O$_{12}$. (a) and (b) displays the decomposed charge densities of Zn$_4$B$_6$O$_{13}$'s highest valence band and lowest conduction band, whereas (c) and (d) plot its band dispersion and DOS at the extended range. As a comparison, (e)-(h) show the isosurface of the decomposed charge density of Zn$_4$B$_6$O$_{12}$'s highest valence band, lowest conduction band, the full band structure and DOS plots. The isosurface value is set as 0.003 e/bohr$^3$ in (a), (b), (e) and (f).}
\end{figure*}

As shown in Fig. \ref{Fig6}, Zn$_4$B$_6$O$_{13}$ is a semiconductor with a band gap of 3.9 eV. Its valence bands near the Fermi level are contributed by Zn and non-cage O atoms. Following the removal of cage O atoms, we observe a lowering of several high-energy conduction bands, similar to what is found in other systems. According to Fig. \ref{Fig6}f, these bands correspond to the antibonding states with localized electrons around the cage. However, we do not find significant electron density around the cage in the top valence bands.
Instead, the HVB of Zn$_4$B$_6$O$_{13}$ in Zn$_4$B$_6$O$_{12}$ mainly consist of electrons around non-cage O and Zn atoms (see Fig. \ref{Fig6}e). This suggests that the majority of excess electrons tend to return to the Zn atoms rather than remaining localized around the cage. Consequently, the band gap of Zn$_4$B$_6$O$_{12}$ does not vary significantly from that of the parent structure.

\begin{figure*}[htbp]
\centering
\includegraphics[width=0.95 \textwidth]{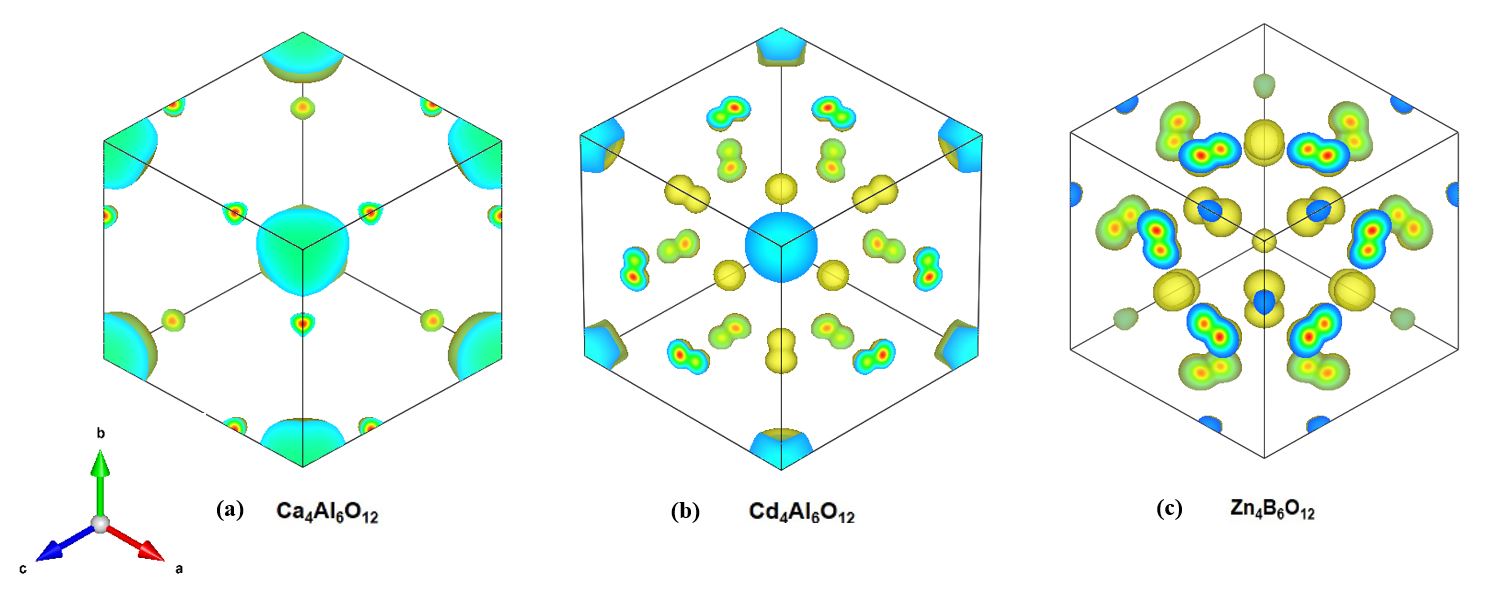}
\caption{\label{Fig7} The comparison of charge redistributions after the removal of cage atoms in the (a) Ca$_4$Al$_6$O$_{12}$, (b) Cd$_4$Al$_6$O$_{12}$ and (c) Zn$_4$B$_6$O$_{12}$. In order to track the charge density near the atomic nucleus, the atomic spheres were intentionally omitted for clarity. The isosurface value is set as 0.004 e/bohr$^3$ to remove some very small charge densities.}
\end{figure*}

Finally, we present the partial charge distribution in the top valence bands of Ca$_4$Al$_6$O$_{12}$, Cd$_4$Al$_6$O$_{12}$, Zn$_4$B$_6$O$_{12}$ in Fig. \ref{Fig7}, where atomic spheres in the unit cell have been removed for clarity. These plots illustrate how the two excess electrons are redistributed following the removal of the cage atoms. In Ca$_4$Al$_6$O$_{12}$, the high valence band (HVB) charge density indicates that the excess electrons remain localized within the cage. However, this arrangement is energetically unfavorable, leading to an upward shift of the HVB in its band structure and a consequent reduction of the band gap. Upon replacing Ca with Cd, the electronic behavior in Cd$_4$Al$_6$O$_{12}$ changes significantly. Due to Cd’s higher electronegativity, a portion of the cage electrons is redistributed towards the Cd nucleus, as shown in Fig. \ref{Fig7}b. In Zn$_4$B$_6$O$_{12}$, the charge redistribution is even more pronounced (see Fig. \ref{Fig7}c). Here, the majority of excess electrons accumulate around the Zn nucleus due to the attractive forces from both neighboring Zn and B nuclei. In both Zn$_4$B$_6$O$_{12}$ and Cd$_4$Al$_6$O$_{12}$, this charge redistribution contributes to maintaining a low energy state, resulting in no significant band gap reduction in these systems.

From the above analysis, we find that after removing the cage center atoms in the sodalite structures, all systems may accommodate electrons at the cage center. However, these electrons may not be sufficient to generate a fully occupied energy band near the Fermi level. The electronegativity of the cations adjacent to the cage plays a crucial role in this phenomenon. Among the four systems studied, only the cage electrons surrounded by the electron-rich Ca$^{2+}$ cations form a fully occupied valence band near the Fermi level. These electrons create an anionic electronic sphere, which interacts weakly with the neighboring Ca$^{2+}$ cations. In contrast, the cage electrons in the other systems may be redistributed to the neighboring Cd$^{2+}$ or Zn$^{2+}$ cations, preventing them from being considered as nucleus-free anions. This observation aligns with our previous screening work \cite{ZHU20191293}, which indicated that only group I, II, and early transition metals can form electrides.

\section{Conclusions}
In this work, we conducted a survey to design new semiconducting electrides by selectively removing anions from existing sodalite structures. Our simulations reveal notable electron localization near the cage center after the removal of anions occupying the high-symmetry Wyckoff sites in the sodalite structures. However, the localized cage electrons may or may not form a distinct electride state in the electronic band structure, primarily depending on the electronegativity of the surrounding cations near the sodalite cages.
Among the candidate compounds, Ca$_4$Al$_6$O$_{12}$ serves as an ideal example of electron localization, capable of forming a fully occupied energy band near the Fermi level. This leads to a semiconducting electride with a significantly reduced band gap of 1.2 eV, compared to the parental sodalite structure’s band gap of 3.9 eV. Additionally, this phase may exhibit improved thermal stability due to its complex structural framework. We hope that our findings will guide further experimental designs for new semiconducting electrides.

\section*{Acknowledgments}
This research was sponsored by the U.S. Department of Energy, Office of Science, Office of Basic Energy Sciences and the Established Program to Stimulate Competitive Research (EPSCoR) under the DOE Early Career Award No. DE-SC0021970. The computing resources are provided by ACCESS (TG-DMR180040).

\bibliography{ref.bib}


\end{document}